\documentclass[pra,twocolumn,superscriptaddress,10pt]{revtex4-1}
\usepackage{graphicx}
\usepackage{dcolumn}
\usepackage{color}
\usepackage{times}
\usepackage{bm}
\usepackage{amssymb}
\usepackage{amsmath}
\usepackage{epsfig}
\usepackage{epstopdf}
\usepackage{dsfont}
\usepackage{subfigure}
\begin{document}
	
	\title{\textbf{Experimental certification of steering criterion based on general entropic uncertainty relation}}
	
	\author{Huan Yang}
	\affiliation{School of Physics and Material Science, Anhui University, Hefei 230601, China}
	\affiliation{Institutes of Physical Science and Information Technology, Anhui University, Hefei 230601, China}
	\affiliation{Department of Experiment and Practical Training Management, West Anhui University, Lu’an 237012, China}
	
	\author{Zhi-Yong Ding}
	\affiliation{School of Physics and Material Science, Anhui University, Hefei 230601, China}
	\affiliation{School of Physics and Electronic Engineering, Fuyang Normal University, Fuyang 236037, China}
	\affiliation{Key Laboratory of Functional Materials and Devices for Informatics of Anhui Educational Institutions, Fuyang Normal University, Fuyang 236037, China}
	
	\author{Dong Wang}
	\affiliation{School of Physics and Material Science, Anhui University, Hefei 230601, China}
	\affiliation{CAS Key Laboratory of Quantum Information, University of Science and Technology of China, Hefei 230026, China}	
	
	\author{Hao Yuan}
	\affiliation{School of Physics and Material Science, Anhui University, Hefei 230601, China}
	\affiliation{CAS Key Laboratory of Quantum Information, University of Science and Technology of China, Hefei 230026, China}	
	\affiliation{Key Laboratory of Opto-Electronic Information Acquisition and Manipulation of Ministry of Education, Anhui University, Hefei 230601, China}
	
	\author{Xue-Ke Song}
	\affiliation{School of Physics and Material Science, Anhui University, Hefei 230601, China}
	
	\author{Jie Yang}
	\affiliation{School of Physics and Material Science, Anhui University, Hefei 230601, China}
	
	\author{Chang-Jin Zhang}
	\affiliation{Institutes of Physical Science and Information Technology, Anhui University, Hefei 230601, China}
	
	\author{Liu Ye}
	\email[]{yeliu@ahu.edu.cn}
	\affiliation{School of Physics and Material Science, Anhui University, Hefei 230601, China}
	
	\begin{abstract}
		Quantum steering describes the phenomenon that one system can be immediately influenced by another with local measurements. It can be detected by the violation of a powerful and useful steering criterion from general entropic uncertainty relation. This criterion, in principle, can be evaluated straightforwardly and achieved by only probability distributions from a finite set of measurement settings. Herein, we experimentally verify the steering criterion by means of the two-photon Werner-like states and three Pauli measurements. The results indicate that quantum steering can be verified by the criterion in a convenient way. In particular, it is no need to perform the usual quantum state tomography in experiment, which reduces the required experimental resources greatly. Moreover, we demonstrate that the criterion is stronger than the linear one for the detecting quantum steering of the Werner-like states.
	\end{abstract}
	
	\maketitle
	
	\section{INTRODUCTION}
		Quantum correlations describe a distinctive phenomenon that they possess stronger correlations between distant subsystems of quantum world than classical one. In principle, quantum correlations can be divided into the three categories: quantum entanglement \cite{w01, w02}, quantum steering \cite{w03}, and Bell nonlocality \cite{w04, w05}, which form a strict hierarchical relationship with each other \cite{w06,w07,w08}, taking a vital part in modern quantum physics \cite{w09, w10}. In particular, quantum steering, coined in the early days of quantum mechanics by Schr$\ddot{\rm o}$dinger, enables one subsystem of an entangled state to influence another by implementing local measurements \cite{w03}. It is easier to be performed than Bell nonlocality \cite{w11}, and has many advantages in applications \cite{w12}. For example, subchannel discrimination \cite{w13}, quantum communication \cite{w14, w15}, randomness generation \cite{w16, w17}, and so on. The investigations concerning quantum steering have been attracted much attentions in the aspects of both theory \cite{w18,w19,w20,w21,w22,w23,w24} and experiment \cite{w25, w26}.
		
		The detection of quantum steering is one of hot topics in quantum information technology, and an effective way to detect quantum steering is to formulate an appropriate criterion for the correlations between the measurement statistics of subsystems \cite{w12}. There are various steering criteria that quantum steering can be verified by their violation, for instance, linear steering criterion (LSC) \cite{w07,w18}, steering criterion from geometric Bell-like inequality \cite{w27}, moment matrix approach \cite{w28,w29}, and steering criteria from entropic uncertainty relations (SCEs) \cite{w30,w31,w32}. Among these, the SCE is extensively investigated owing to their conceptual simplicity \cite{w32}, and it also can reveal the connection between steering and entropic uncertainty relations.
		
		Theoretically, in 2011, Walborn \textit{et al}. \cite{w30} proposed the SCE for the bipartite continuous variable systems, and provided a way to witness quantum steering. Subsequently, Schneeloch \textit{et al}. \cite{w31} extended the SCE to the discrete systems. The SCE for multipartite systems was derived by Riccardi \textit{et al}. \cite{w35}, and different classes of multipartite steering were certified by the violation of their SCE. All of these theoretical efforts are based on Shannon entropies. Recently, by using Tsallis entropies instead of Shannon ones, Costa \textit{et al}. \cite{w36} derived steering criterion from general entropic uncertainty relation (SCG). Typically, the criterion is stronger than the ones from Shannon entropies and LSC \cite{w36}. Moreover, it is no need to use semidefinite programming.
		
		On the experimental side, the tests of steering criteria have been demonstrated in several promising experiments \cite{w37,w38,w39,w40,w41,w42}. Saunders \textit{et al}. experimentally observed quantum steering for Bell local state via LSC \cite{w37}. In 2018, Li \textit{et al}. \cite{w41} verified steering criterion from geometric Bell-like inequality by performing quantum state tomography in experiment. In 2019, by utilizing fine-grained steering criterion \cite{w43}, Pramanik \textit{et al}. \cite{w42} revealed hidden quantum steerability with the help of local filtering operations. However, the exploration of the SCG remains in theory, and the experimental investigation concerning it is still lacking. In addition, the SCG is easy to be implemented due to that it directly depends on probability distributions from a finite set of measurements \cite{w36}. As a consequence, the experimental operation may be simplified via the criterion, and it also provides a convenient and effective tool for us to capture quantum steering in practice.
		
		In this work, we experimentally verify the SCG by using an all-optical setup. It shows that our experimental results agree very well with theoretical predictions. According to probability distributions of a few measurements in the experiment, one can directly and conveniently detect quantum steering of bipartite states without implementing tomography on quantum states. Furthermore, we certify that the SCG is stronger than LSC in the detecting steering of the Werner-like states.
			
	\section{STEERING CRITERION FROM GENERAL ENTROPIC UNCERTAINTY RELATION}
		Assuming Alice and Bob share a two-qubit state ${\rho _{AB}} = (\mathbb{I} \otimes \mathbb{I} + {\bm{a}} \cdot {\bm{\sigma }} \otimes \mathbb{I} + \mathbb{I} \otimes {\bm{b}} \cdot {\bm{\sigma }} + \sum\nolimits_{i,j = 1}^3 {{c_{ij}}{\sigma _i} \otimes {\sigma _j}} )/4$, where ${\bm{\sigma }}$ denotes the vector of Pauli matrices. ${\bm{a}} = {\rm{tr}}\left( {{\rho _{AB}}{\bm{\sigma }} \otimes \mathbb{I}} \right)$ and ${\bm{b}} = {\rm{tr}}\left( {{\rho _{AB}}\mathbb{I} \otimes {\bm{\sigma }}} \right)$ are the Bloch vectors of Alice's and Bob's reduced states. The matrix element of the spin correlation matrix is ${c_{ij}} = {\rm{tr}}\left( {{\rho _{AB}}{\sigma _i} \otimes {\sigma _j}} \right){\rm{ }}$. Alice implements a measurement \textit{A} on her part of the system, and then obtains the measurement outcome $i$. And, Bob implements a measurement \textit{B} with measurement outcome $j$ on his subsystem. Considering all possible measurements \textit{A} and \textit{B}, the joint probability distribution of the outcomes can be represented by \cite{w06}
		\begin{equation}\label{g01}
			p(i,j|A,B) = \sum\limits_\lambda  {p(\lambda )p(i|A,\lambda ){p_q}(j|B,\lambda )},
		\end{equation}
		where $p(i|A,\lambda )$ indicates a general probability distribution, ${p_q}(j|B,\lambda ) = {\rm{T}}{{\rm{r}}_B}[B(j){\sigma _\lambda }]$ denotes a probability distribution originating from a quantum state ${\sigma _\lambda }$, and $B(j)$ is a measurement operator with $\sum\nolimits_j {B(j)}  = \mathbb{I}$ and $\sum\nolimits_\lambda  {p(\lambda )}  = 1$. The model as in Eq. (1) is known as a local hidden state model. And the state is regarded as being steerable when the measured correlations do not admit the local hidden state model.

		In the next section, let us briefly review the SCG. To begin with, we introduce the Tsallis entropy \cite{w44}, as a possible generalized entropy, which is defined as ${S_q}(P) =- \sum\nolimits_i {p_i^q{\rm{l}}{{\rm{n}}_q}({p_i})}$, here ${\rm{l}}{{\rm{n}}_q}(x) = ({x^{1 - q}} - 1)/(1 - q)$ and parameter $q > 1$. This entropy can recover to the Shannon entropy $S(p)$ for $q \to 1$. And then, we give an example to illustrate the entropic uncertainty relations. One carries out Pauli measurements ${\sigma _x}$ and ${\sigma _z}$ on a single qubit. These measurements generate a two-valued probability distribution and the corresponding entropy, which can be represented by $S({\sigma _x}) + S({\sigma _z}) \ge \ln (2)$ \cite{w45}. For $k$ measurements ${{B_k}}$, and they obey $\sum\nolimits_k {S({B_k})}  \ge \mathcal{B}$, where $\mathcal{B}$ is the bound of entropic uncertainty relation. If one performs a set of measurements ${A_k} \otimes {B_k}$ on the bipartite systems, we have $\sum\nolimits_k {S({B_k}|{A_k})}  \ge {C_B}$ \cite{w31,w36}, where ${S(B|A)}$ is the conditional entropy and $C_B$ is the bound. Any nonsteerable state obeys the inequality. Therefore, the steering criteria can be derived via entropic uncertainty relations. Based on the Tsallis entropy and entropic uncertainty relation, Costa \textit{et al}. derived the SCG as \cite{w36}				
		\begin{equation}\label{g02}
			\sum\limits_k {[{S_q}({B_k}|{A_k}) + (1 - q)C({A_k},{B_k})} ] \ge C_B^{(q)}
		\end{equation}
		with $C(A,B) = \sum\nolimits_i {p_i^q{{[{{\ln }_q}({p_i})]}^2}}  - \sum\nolimits_{i,j} {p_{ij}^q{{\ln }_q}({p_i}){{\ln }_q}({p_{ij}})}$ and ${S_q}(B|A) = {S_q}(A,B) - {S_q}(A)$ \cite{w46}. One can detect steering of a two-qubit state by the violation of the SCG. It is worth emphasizing that the SCG can be rewritten as
		\begin{equation}\label{g03}
			\frac{1}{{q - 1}}\left[ {\sum\limits_k {\left( {1 - \sum\limits_{ij} {\frac{{{{(p_{ij}^{(k)})}^q}}}{{{{(p_i^{(k)})}^{q - 1}}}}} } \right)} } \right] \ge C_B^{(q)},
		\end{equation}
		where $p_{ij}^{(k)}$ is the probability of outcome $(i,j)$ for Alice and Bob by performing measurement set ${A_k} \otimes {B_k}$, and $p_i^{(k)}$ denotes the probability of marginal outcome by implementing measurement  ${A_k}$ on Alice. For the Tsallis entropy, considering $d$-dimension system and $m$ mutually unbiased bases (MUBs), the bounds $C_B^{(q)}$ for $q \in (0;2]$ are given by \cite{w47}
		\begin{equation}\label{g04}
		C_B^{(q)} = m{\ln _q}\left( {\frac{{md}}{{d + m - 1}}} \right).
		\end{equation}
		For even dimensions, this bound is not optimal in the case of $q \to 1$. It is more appropriate to replace the bounds mentioned above with the results in Ref. \cite{w48}
		\begin{equation}\label{g05}
			{C_B} = \left\{ {\begin{array}{*{20}{l}}
			{(d + 1)\ln \left( {\frac{{d + 1}}{2}} \right),}&{d\;\rm{is}\;{\rm{odd}}}\\
			{\frac{d}{2}\ln \left( {\frac{d}{2}} \right) + (\frac{d}{2} + 1)\ln \left( {\frac{d}{2} + 1} \right),}&{d\;\rm{is}\;{\rm{even}}}
			\end{array}} \right.
		\end{equation}
		
		Theoretically, the criterion of Eq. (3) is the strongest one for $q$ = 2 \cite{w36}. As an illustration, by considering a two-qubit Werner state $\rho  = \chi \left| {{\phi ^ + }} \right\rangle \left\langle {{\phi ^ + }} \right| + (1 - \chi )\mathbb{I}/4$, with $\chi  \in [0,1]$ and $\left| {{\phi ^ + }} \right\rangle  = \left( {\left| {00} \right\rangle  + \left| {11} \right\rangle } \right)/\sqrt 2 $, the SCG for  $q = 2$  can be violated when $\chi  > 1/\sqrt 3  \approx 0.577$ with a complete set of MUBs, which is termed as the optimal threshold \cite{w36}. Note that, the SCG directly depends on probability distributions from a finite set of measurements and is straightforward to evaluate. What's more, this criterion does not need to implement full tomography on quantum state, and this criterion is therefore easy to be performed in experiment.
		
	\section{EXPERIMENTAL IMPLEMENTATION AND RESULTS}
		In order to experimentally verify the SCG, a bipartite qubit state with high fidelity is firstly needed to prepare in our photon-polarization-qubit system. And then, we choose measurement settings in the Pauli \textit{X}, \textit{Y} and \textit{Z} bases as three MUBs. To be specific, we use the time-mixing technique \cite{w49,w50,w51} to prepare the two-photon Werner-like states,
		\begin{equation}\label{g06}
			{\rho _{\rm{w}}} = \chi \left| {{\phi _{AB}}} \right\rangle \left\langle {{\phi _{AB}}} \right| + (1 - \chi )\frac{\mathbb{I}}{4}
		\end{equation}
		with $\left| {{\phi _{AB}}} \right\rangle {\rm{ = }}\cos 2\theta \left| {HH} \right\rangle  + \sin 2\theta \left| {VV} \right\rangle $ , where $H$ and $V$ represent the horizontally and vertically polarized components, respectively. Hence, the preparation of the Werner-like states require the mixture of the entangled state $\left| {{\phi _{AB}}} \right\rangle $ with the identity state $\mathbb{I}/4$. The relative weights $\chi $ in the mixture are responsible for through appropriate measurement durations \cite{w49,w52,w53}.
		
		\begin{figure}
			\centering
			\includegraphics[width=8cm]{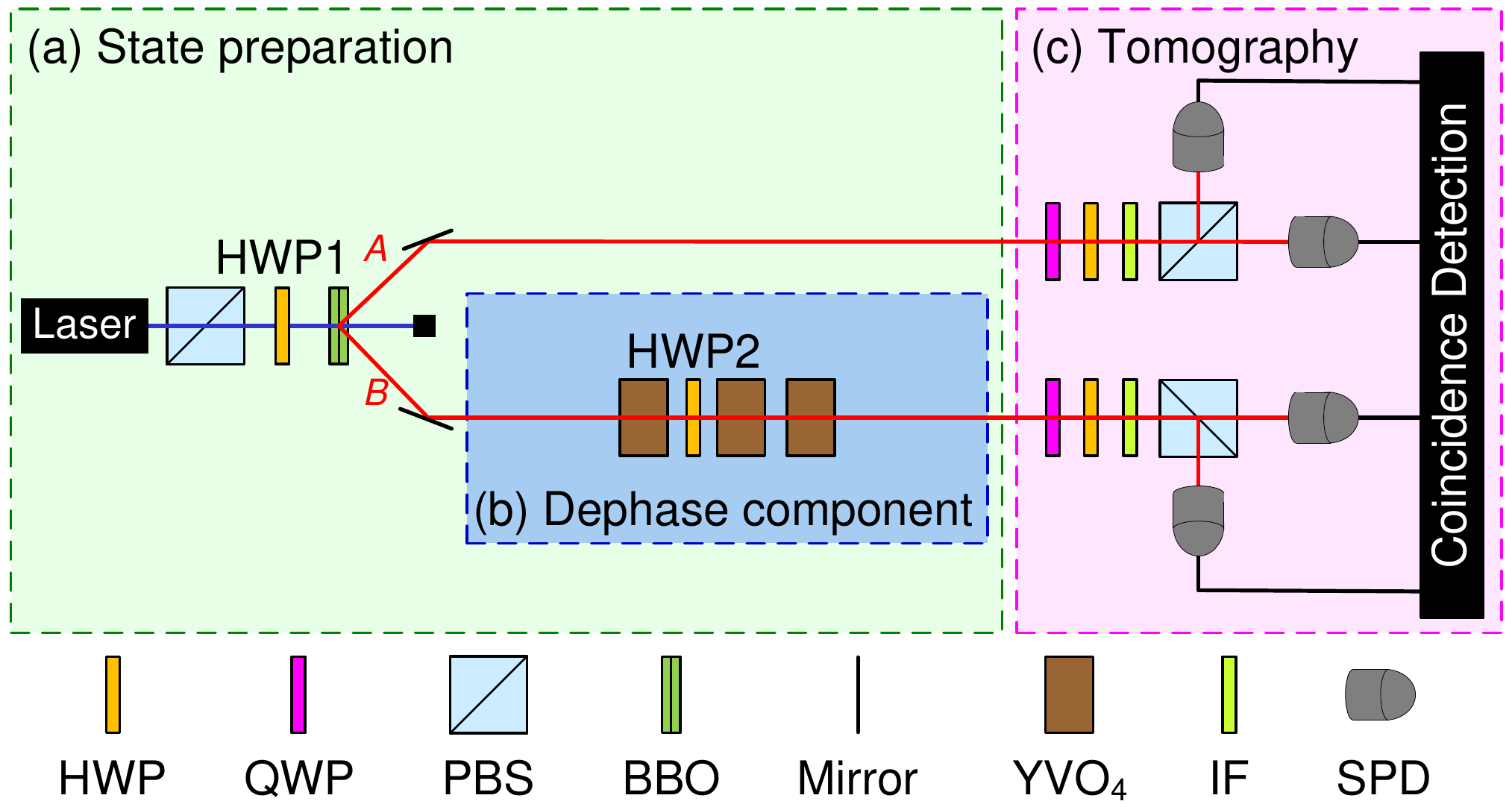}\\
			\caption{Experimental setup is constructed by three modules: (a) State preparation module, (b) dephase component, and (c) tomography module. In module (a), two-photon states $\left| {{\phi _{AB}}} \right\rangle {\rm{ = }}\cos 2\theta \left| {HH} \right\rangle  + \sin 2\theta \left| {VV} \right\rangle $ are prepared by the procession of a spontaneous parametric down-conversion. In module (b), three ${\rm{YV}}{{\rm{O}}_4}$ crystals and a HWP with ${22.5^{\rm{o}}}$ compose dephase component, which can dephase the two-photon state into a completely mixed state $\mathbb{I}/4$. Module (c) is used to realize three Pauli measurements and reconstruct quantum states. Abbreviations: HWP, half-wave plate; QWP, quarter-wave plate; PBS, polarizing beam splitter; BBO, type-I $\beta $-barium borate crystal; ${\rm{YV}}{{\rm{O}}_4}$, yttrium orthovanadate crystal; IF, interference filter; SPD, single photon detector.}\label{Fig3}
		\end{figure}
		
		The schematic diagram of our experimental setup is illustrated in Fig. 1. The setup consists of three modules: the green frame represents state preparation module (a), the blue frame represents dephase component module (b) and the red frame is tomography module (c). In module (a), the horizontally polarized state $\left| H \right\rangle $ is generated by a high-power continuous laser beam (130 mW, 405 nm) passes through the polarization beam splitter (PBS). After the transmitted beam passes through the half-wave plate 1 (HWP1) and two type-I $\beta $-barium borate (BBO) crystals (${6.0 \times 6.0 \times 0.5}$ mm), a pair of entangled photons state $\left| {{\phi _{AB}}} \right\rangle {\rm{ = }}\cos 2\theta \left| {HH} \right\rangle  + \sin 2\theta \left| {VV} \right\rangle $  ($\lambda $= 810 nm) can be obtained via the spontaneous parametric down-conversion \cite{w54}. The ratio of $\left| {HH} \right\rangle $ to $\left| {VV} \right\rangle $ can be adjusted via modulating the angle of optical axis of HWP1. The module (b) is used to generate the identity state $\mathbb{I}/4$. Firstly, the angle of optical axis of HWP1 is set to ${22.5^{\rm{o}}}$, and the maximum entangled state $(\left| {HH} \right\rangle  + \left| {VV} \right\rangle )/\sqrt 2 $ is prepared. And then, let the photon of \textit{B}-path through the module (b), which is composed by three 2.6 mm yttrium orthovanadate $({\rm{YV}}{{\rm{O}}_4})$ crystals and a HWP2 with ${22.5^{\rm{o}}}$ \cite{w55}. This combination can dephase the two-photon state into a completely mixed state $\mathbb{I}/4$. The first role of module (c) is to realize three local Pauli measurements for both photons \textit{A} and \textit{B}. By rotating the angles of HWP and QWP in the module (c), we can obtain the six measurement settings, as shown in Table I. For each measurement, the outcome is marked as 0 or 1. The second role of module (c) is to reconstruct quantum states by using tomography \cite{w56}, and the purpose is to obtain the fidelity of prepared states rather than the verification of SCG.

		 \begin{table}[b]
			\centering
			\caption{Types and angles of wave plates for different settings of local Pauli measurement.}
			\begin{ruledtabular}
				\begin{tabular}{ccccc}
					Measurement & \rm{The angle of HWP} & \rm{The angle of QWP} \\
					\colrule
					$\Pi _x^0$ &  $22.5 ^{\circ}$ & $45 ^{\circ}$\\
					$\Pi _x^1$ & $-22.5 ^{\circ}$ &  $45 ^{\circ}$\\
					$\Pi _y^0$ & $22.5 ^{\circ}$ &  $0 ^{\circ}$\\
					$\Pi _y^1$ &  $-22.5 ^{\circ}$ &  $0 ^{\circ}$\\
					$\Pi _z^0$ &  $0 ^{\circ}$ &  $0 ^{\circ}$\\
					$\Pi _z^1$ & $45 ^{\circ}$ &  $0 ^{\circ}$\\
				\end{tabular}
			\end{ruledtabular}		
		\end{table}

		\begin{figure}
			\centering
			\includegraphics[width=8cm]{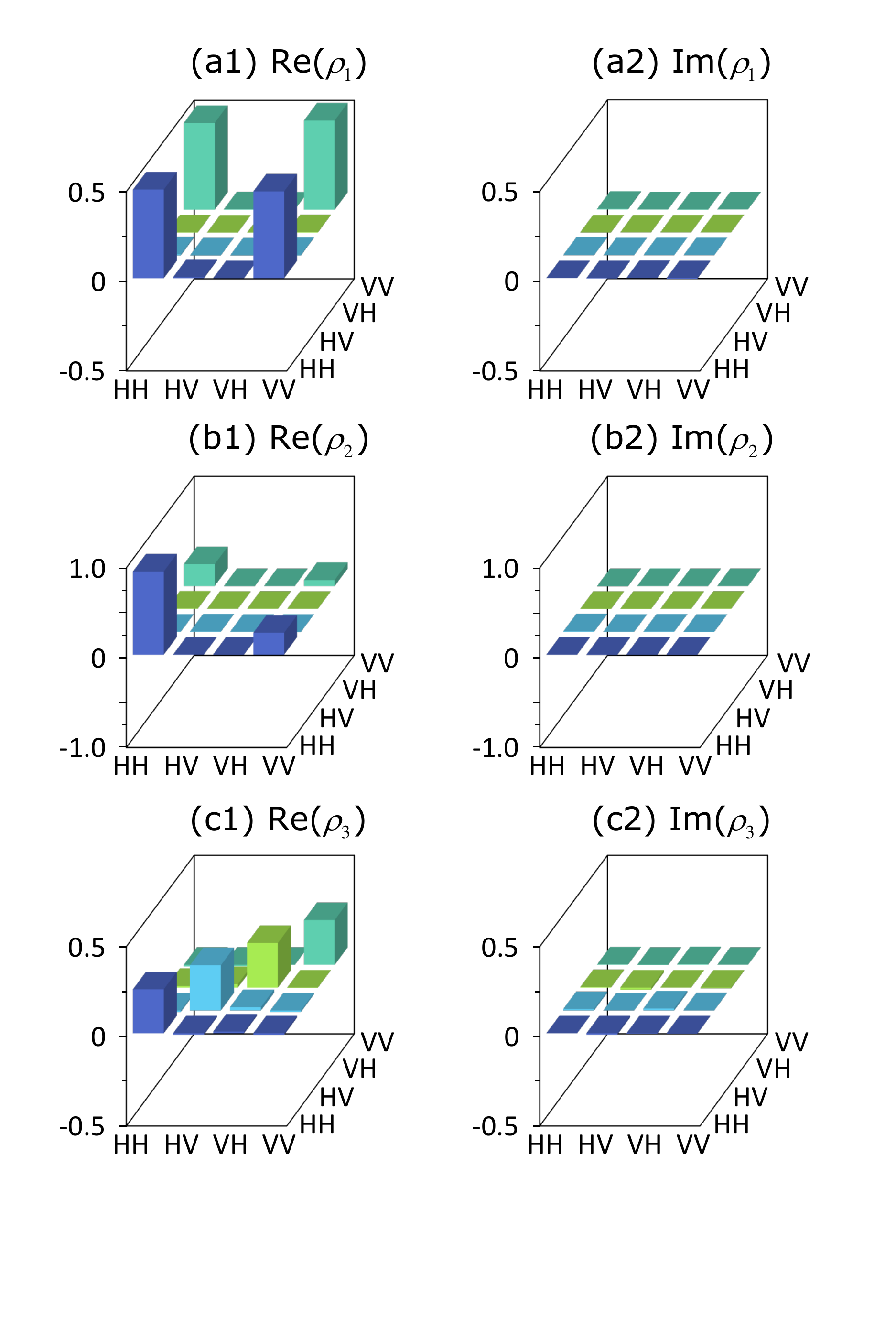}\\
			\caption{The reconstructed density matrices of ${\rho _1}(\theta {\rm{ = }}{22.5^{\rm{o}}}, \chi {\rm{ = }}1)$,  ${\rho _2}(\theta {\rm{ = }}{7.5^{\rm{o}}}, \chi {\rm{ = }}1)$ and ${\rho _3}(\chi {\rm{ = }}0)$. ${\mathop{\rm Re}\nolimits} ( \cdot )$ and ${\mathop{\rm Im}\nolimits} ( \cdot )$  represent the real and imaginary parts of these states, respectively.}\label{Fig3}
		\end{figure}

	  	\begin{table}[b]
	  		\centering
	  		\caption{Fidelity of the prepared states with different state parameters $\theta$ and $\chi$.}
	  		\begin{ruledtabular}
	  			\begin{tabular}{ccccc}
	  				\multicolumn{2}{c}{$\theta {\rm{ = }}{22.5^{\rm{o}}}$} & \multicolumn{2}{c}{$\theta {\rm{ = }}{7.5^{\rm{o}}}$} \\
	  				\colrule
	  				$\chi$ = 0.00 & 0.9991$\pm $0.0002 & $\chi$ = 0.00 & 0.9991$\pm $0.0002 \\
	  				$\chi$ = 0.10 & 0.9977$\pm $0.0004 & $\chi$ = 0.15 & 0.9979$\pm $0.0004 \\
	  				$\chi$ = 0.34 & 0.9998$\pm $0.0001 & $\chi$ = 0.30 & 0.9998$\pm $0.0001 \\
	  				$\chi$ = 0.42 & 0.9998$\pm $0.0003 & $\chi$ = 0.45 & 0.9972$\pm $0.0005 \\
	  				$\chi$ = 0.50 & 0.9998$\pm $0.0001 & $\chi$ = 0.55 & 0.9995$\pm $0.0002 \\
	  				$\chi$ = 0.58 & 0.9998$\pm $0.0002 & $\chi$ = 0.65 & 0.9997$\pm $0.0002 \\
	  				$\chi$ = 0.66 & 0.9997$\pm $0.0002 & $\chi$ = 0.75 & 0.9994$\pm $0.0003 \\
	  				$\chi$ = 0.74 & 0.9998$\pm $0.0003 & $\chi$ = 0.81 & 0.9994$\pm $0.0004 \\
	  				$\chi$ = 0.90 & 0.9989$\pm $0.0010 & $\chi$ = 0.89 & 0.9995$\pm $0.0004 \\
	  				$\chi$ = 1.00 & 0.9929$\pm $0.0019 & $\chi$ = 1.00 & 0.9968$\pm $0.0006 \\
	  			\end{tabular}
	  		\end{ruledtabular}		
	  	\end{table}

		In the experimental verification, we focus on two classes of states. The first one is Werner states, which satisfy ${\bm{a}} = {\bm{b}} = 0$. Experimentally, the state parameter $\theta $	in Eq. (6) is set to ${22.5^{\rm{o}}}$, and adjust mixture weights $\chi$ ($\chi$ = 0.00, 0.10, 0.34, 0.42, 0.50, 0.58, 0.66, 0.74, 0.90, 1.00). The average fidelity of these states is $\bar F = 0.9987 \pm 0.0005$. (the fidelity of each state is shown in Table II). All error bars in the experiment are evaluated based on the standard deviation from the statistical variation of the photon counts, which are assumed to follow Poisson distribution. The second class of test states is Werner-like states, which satisfy ${\bm{a}} = {\bm{b}} \ne 0$. We choose $\theta {\rm{ = }}{7.5^{\rm{o}}}$ and adjust $\chi$ ($\chi$ = 0.00, 0.15, 0.30, 0.45, 0.55, 0.65, 0.75, 0.81, 0.89, 1.00). As a result, various initial Werner-like states can be prepared. The average fidelity of these states is $\bar F = 0.9988 \pm 0.0003$ (the fidelity of each state is shown in Table II). Fig. 2 provides the real and imaginary parts of three reconstructed states: the maximally entangled states ${\rho _1} (\theta {\rm{ = }}{22.5^{\rm{o}}}, \chi {\rm{ = }}1)$, entangled states ${\rho _2} (\theta {\rm{ = }}{7.5^{\rm{o}}}, \chi {\rm{ = }}1)$, and the identity state ${\rho _3} (\chi {\rm{ = }}0)$. It shows that these states have high fidelity. The Supplemental Material offers the explicit tomographic results for all the states reconstructed. With the help of these high-fidelity states, we can experimentally test the SCG. Remarkably, the probabilities $p_{ij}^{(k)} (k = x,y,z\;{\rm{and }}\;i,j = 0,1)$ and $p_i^{(k)}$ in Eq. (3) can be conveniently and directly calculated via the coincidence counts in experiment (detailed outcomes see Tables 1 and 2 in Supplemental Material), and it does not need to be calculated according to the reconstructed quantum states. Hence, the values of the left-hand side (LHS) of the SCG are numerically calculated by utilizing these probabilities.
		
		In addition, LSC \cite{w18} is another widely used steering criterion.  Considering three measurement settings, there have a concise inequality ${(\sum\nolimits_{i = 1}^3 {c_i^2} )^{1/2}} \le 1$ for any two-qubit state ${\rho _{AB}} = (\mathbb{I} \otimes \mathbb{I} + {\bm{a}} \cdot {\bm{\sigma }} \otimes \mathbb{I} + \mathbb{I} \otimes {\bm{b}} \cdot {\bm{\sigma }} + \sum\nolimits_{i = 1}^3 {{c_i}{\sigma _i} \otimes {\sigma _i}} )/4$, where ${c_i} = {\rm{tr}}\left( {{\rho _{AB}}{\sigma _i} \otimes {\sigma _i}} \right){\rm{ }}$. The violation of the LSC implies steerability of the state. Considering three Pauli measurements and $q = 2$, the SCG is stronger than LSC for the detecting steering of any two-qubit state due to the SCG uses more information ($\bm{a}$ and $\bm{b}$) about the state \cite{w36}. If ${\bm{a}} = {\bm{b}} = 0$, the two criteria are equivalent \cite{w36}. In our experiment, we obtain the values of ${c_i}$ via the density matrix reconstructed of states based on tomography process, and compare the SCG with LSC for the detection of steerable states using three measurements.
		
		\begin{figure}
			\centering
			\includegraphics[width=8cm]{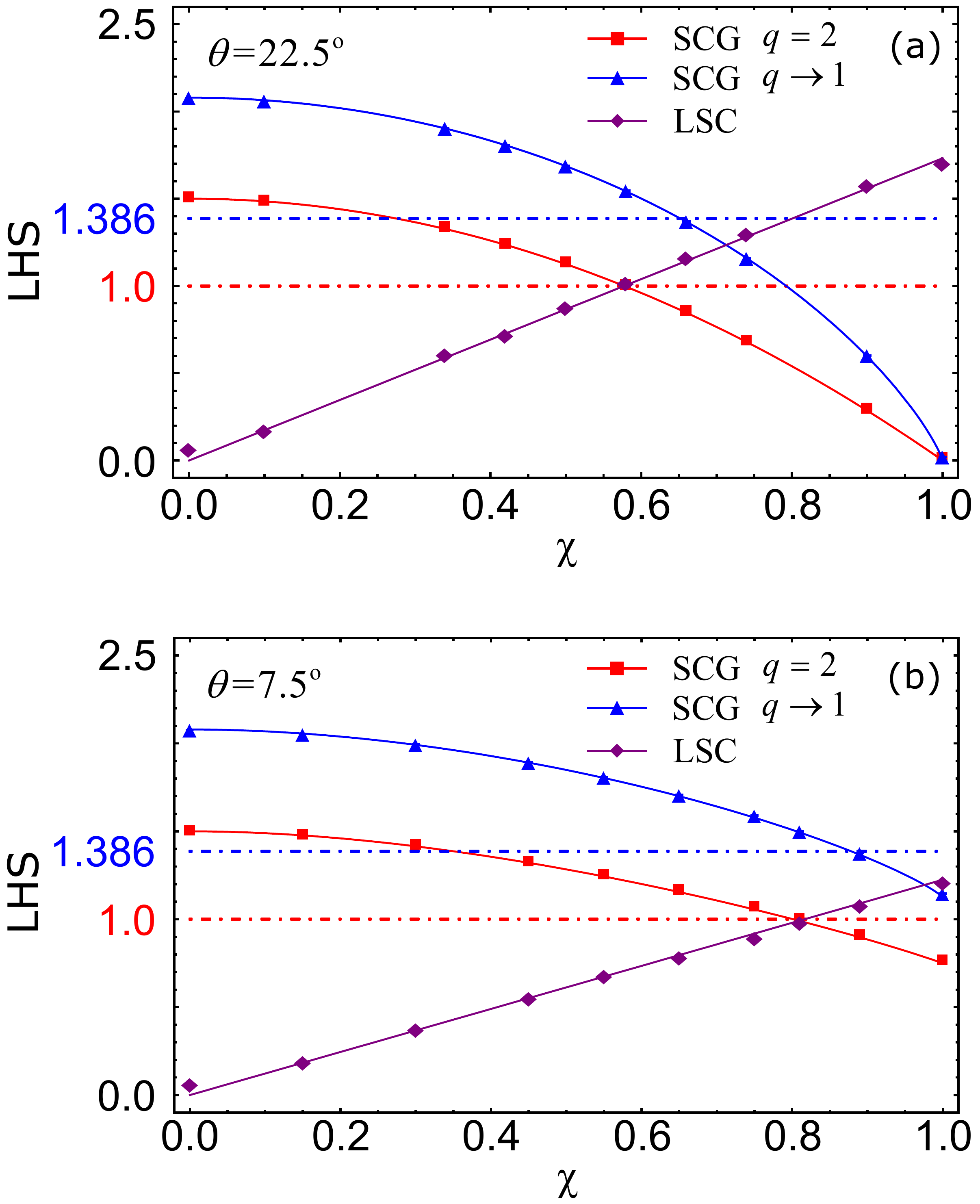}\\
			\caption{Experimental results and the corresponding theoretical predictions of LHS for the SCG and LSC. The red horizontal dot dash line represents the bounds of the SCG for $q = 2$ and LSC. The blue horizontal dot dash line represents the bound of the SCG for  $q \to 1$.}\label{Fig3}
		\end{figure}

		\begin{table}[b]
			\centering
			\caption{Detailed experimental datas of LHS of the SCG ($q$ = 2 and $q\to 1$) and LSC for different Werner states $(\theta {\rm{ = }}{22.5^{\rm{o}}})$.}
			\begin{ruledtabular}
				\begin{tabular}{ccccc}
					$\theta {\rm{ = }}{22.5^{\rm{o}}}$ & SCG for $q$ = 2 & SCG for $q\to 1$ & LSC\\
					\colrule
					$\chi$ = 0.00 & 1.4993$\pm $0.0001 & 2.0787$\pm $0.0001 & 0.0608$\pm $0.0029 \\
					$\chi$ = 0.10 & 1.4807$\pm $0.0006 & 2.0601$\pm $0.0006 & 0.1668$\pm $0.0029 \\
					$\chi$ = 0.34 & 1.3279$\pm $0.0017 & 1.9038$\pm $0.0017 & 0.6028$\pm $0.0030 \\
					$\chi$ = 0.42 & 1.2340$\pm $0.0021 & 1.8049$\pm $0.0023 & 0.7141$\pm $0.0029 \\
					$\chi$ = 0.50 & 1.1255$\pm $0.0025 & 1.6875$\pm $0.0028 & 0.8729$\pm $0.0029 \\
					$\chi$ = 0.58 & 0.9982$\pm $0.0028 & 1.5450$\pm $0.0032 & 1.0133$\pm $0.0027 \\
					$\chi$ = 0.66 & 0.8471$\pm $0.0027 & 1.3683$\pm $0.0032 & 1.1582$\pm $0.0023 \\
					$\chi$ = 0.74 & 0.6781$\pm $0.0028 & 1.1585$\pm $0.0035 & 1.2945$\pm $0.0022 \\
					$\chi$ = 0.90 & 0.2883$\pm $0.0022 & 0.6009$\pm $0.0036 & 1.5728$\pm $0.0015 \\
					$\chi$ = 1.00 & 0.0048$\pm $0.0003 & 0.0195$\pm $0.0009 & 1.6995$\pm $0.0008 \\
				\end{tabular}
			\end{ruledtabular}		
		\end{table}
	
			\begin{table}[b]
			\centering
			\caption{Detailed experimental datas of LHS of the SCG ($q$ = 2 and $q\to 1$) and LSC for different Werner-like states $(\theta {\rm{ = }}{7.5^{\rm{o}}})$.}
			\begin{ruledtabular}
				\begin{tabular}{ccccc}
					$\theta {\rm{ = }}{7.5^{\rm{o}}}$ & SCG for $q$ = 2 & SCG for $q\to 1$ & LSC\\
					\colrule
					$\chi$ = 0.00 & 1.4993$\pm $0.0001 & 2.0787$\pm $0.0001 & 0.0608$\pm $0.0029 \\
					$\chi$ = 0.15 & 1.4747$\pm $0.0007 & 2.0539$\pm $0.0007 & 0.1865$\pm $0.0031 \\
					$\chi$ = 0.30 & 1.4179$\pm $0.0013 & 1.9953$\pm $0.0013 & 0.3727$\pm $0.0031 \\
					$\chi$ = 0.45 & 1.3236$\pm $0.0016 & 1.8935$\pm $0.0018 & 0.5504$\pm $0.0029 \\
					$\chi$ = 0.55 & 1.2484$\pm $0.0017 & 1.8096$\pm $0.0019 & 0.6772$\pm $0.0026 \\
					$\chi$ = 0.65 & 1.1611$\pm $0.0021 & 1.7071$\pm $0.0025 & 0.7833$\pm $0.0027 \\
					$\chi$ = 0.75 & 1.0664$\pm $0.0022 & 1.5912$\pm $0.0027 & 0.8931$\pm $0.0025 \\
					$\chi$ = 0.81 & 0.9965$\pm $0.0021 & 1.5012$\pm $0.0026 & 0.9810$\pm $0.0024 \\
					$\chi$ = 0.89 & 0.9054$\pm $0.0025 & 1.3775$\pm $0.0032 & 1.0786$\pm $0.0025 \\
					$\chi$ = 1.00 & 0.7605$\pm $0.0019 & 1.1470$\pm $0.0022 & 1.2092$\pm $0.0019 \\
				\end{tabular}
			\end{ruledtabular}		
		\end{table}
		
		The verification results of the experiment are shown in Fig. 3, Tables III and IV. Illustratively, we focus on the case of $q = 2$ and $q \to 1$ in the SCG, since the former is the optimal value of $q$ for the detection steering via the SCG and the latter has to do with the usual entropic steering criteria \cite{w36}. In Fig. 3, the horizontal axis represents the parameter $\chi$ of the states in Eq. (6). The red squares and blue triangles represent the experimental LHSs of the SCG for $q = 2$ and $q \to 1$, respectively. The purple rhombuses represent the experimental LHSs of LSC. The solid lines in different colors are the corresponding theoretical predictions. The red horizontal dot dash line represents the bounds of the SCG for $q = 2$ and LSC, which is equal to one. The state is steerable below this line according to the SCG for $q = 2$, and is steerable above this line according to LSC. The blue horizontal dot dash line represents the bound of the SCG for  $q \to 1$, which is approximately equal to 1.386. The state is steerable below this line according to the SCG for $q \to 1$. The detailed experimental datas of LHS are given in Tables III and IV.	It is worth mentioning that the error bars are very small and not displayed in the figures. As seen in Fig. 3, our experimental results coincide with the theoretical predictions very well. From these experimental results, we can conclude:
		
		(1) One can find from Fig. 3(a) and Table III that the SCG for $q = 2$ can be violated (i.e., the LHS of SCG for $q = 2$ is less than one) when $\chi  > 1/\sqrt 3  \approx 0.58$. Five prepared Werner states ($\theta {\rm{ = }}{22.5^{\rm{o}}}$ and $\chi$ = 0.58, 0.66, 0.74, 0.90, 1.00) are verified to be steerable states by SCG for $q = 2$, which means that the Werner states is steerable for $\chi  > 1/\sqrt 3 $ and unsteerable for $\chi  \le 1/\sqrt 3$. The SCG for $q \to 1$ is violated  (i.e., the LHS of SCG for $q \to 1$ is less than 1.386) when $\chi  > 0.652 \approx 0.66$. And only four steerable Werner states ($\theta {\rm{ = }}{22.5^{\rm{o}}}$ and $\chi$ = 0.66, 0.74, 0.90, 1.00) are certified by SCG for $q \to 1$. The results verify that the quantum steering can be conveniently detected via SCG, which is the strongest one for $q = 2$.
		
		(2) As shown in Fig. 3(a) and Table III, the prepared Werner states with $\chi  > 1/\sqrt 3  \approx 0.58$ can also violate the LSC (namely the LHS of LSC is larger than one). This experimental result is consistent with that of SCG for $q = 2$. It certifies that the SCG for $q = 2$ and LSC are equivalent in the detecting steering of the Werner states which satisfy ${\bm{a}} = {\bm{b}} = 0$.
		
		(3) As illustrated in Fig. 3(b) and Table IV, the SCG for $q = 2$  can be violated for the Werner-like states $(\theta {\rm{ = }}{7.5^{\rm{o}}})$ if $\chi  \ge 0.81$. The Werner-like states $(\theta {\rm{ = }}{7.5^{\rm{o}}})$ with $\chi  \ge 0.89$ can violate both the SCG for $q \to 1$ and the LSC. Notably, the Werner-like states with  $\chi$ = 0.81 violate only the SCG for $q = 2$, and cannot violate the SCG for $q \to 1$ and the LSC. The results certify that the SCG for  $q = 2$ is stronger than LSC in the case of ${\bm{a}} = {\bm{b}} \ne 0$, which demonstrates that the SCG contains more information about the state and can detect weakly steerable states.
			
	\section{CONCLUSIONS}	
		In conclusion, by preparing two-photon Werner-like states with high fidelity via time-mixing technique, we experimentally certify the steering criterion from general entropic uncertainty relation. The experimental results show good agreement with theoretical results. Without the implementation of quantum states tomography process, one can directly and effectively detect quantum steering by using probability distributions from few measurements in the experiment. The experimental results also verify that the steering criterion from general entropic uncertainty relation is the strongest one with respect to $q = 2$, which can detect weakly steerable states. Moreover, we confirm that the criterion for $q = 2$ is equivalent to linear steering criterion in the detecting steering of the Werner states. However, the criterion for $q = 2$ is stronger than the linear one in the scenario of Werner-like states due to that it uses more information about the states. Thus, the steering criterion from general entropic uncertainty relation is demonstrated as a convenient tool to detect quantum steering in experiment.
		
	\section*{ACKNOWLEDGMENTS}	
		This work was supported by the National Science Foundation of China (Grant Nos. 11575001, 11405171, 61601002 and 11605028), the Program for Excellent Talents in University of Anhui Province of China (Grant No. gxyq2018059 and gxyqZD2019042), the Natural Science Research Project of Education Department of Anhui Province of China (Grant No. KJ2018A0343), the Open Foundation for CAS Key Laboratory of Quantum Information under Grant Nos. KQI201801 and KQI201804, the Key Program of Excellent Youth Talent Project of the Education Department of Anhui Province of China under Grant No. gxyqZD2018065.
		
		Huan Yang and Zhi-Yong Ding contributed equally to this work.

	\bibliographystyle{plain}
	
\end{document}